%% file: main.tex
\documentclass{article}

\usepackage[preprint]{spconf}
\usepackage[letterpaper,top=2cm,bottom=2cm,left=3cm,right=3cm,marginparwidth=1.75cm]{geometry}
\usepackage{IEEEtrantools}
\usepackage{amsmath}
\usepackage{amssymb}
\usepackage{algorithm}
\usepackage[noend]{algpseudocode}
\usepackage{graphicx}
\usepackage[nolist,nohyperlinks]{acronym}
\usepackage[colorlinks=true, allcolors=blue]{hyperref}
\usepackage{cite}

\algnewcommand\algorithmicforeach{\textbf{for each}}
\algdef{S}[FOR]{ForEach}[1]{\algorithmicforeach\ #1\ \algorithmicdo}
\newcommand{\pluseq}{\mathrel{+}=}

\begin{acronym}
    \input{sapacronyms.txt}
\end{acronym}

\newlength{\bibitemsep}
\newlength{\bibparskip}
\let\oldthebibliography\thebibliography
\renewcommand\thebibliography[1]{%
\oldthebibliography{#1}%
\setlength{\parskip}{\bibitemsep}%
\setlength{\itemsep}{\bibparskip}%
}

\title{Graph neural networks for sound source localization on distributed microphone networks}
\name{Eric Grinstein \sthanks{Contact: e.grinstein@imperial.ac.uk} \sthanks{This project has received funding from the European Union’s Horizon 2020 research and innovation programme under the Marie Skłodowska-Curie grant agreement No 956369}, Mike Brookes, Patrick A. Naylor}

\begin{document}
\maketitle

\begin{abstract}
\acfp{DMA} present many challenges with respect to centralized microphone arrays. An important requirement of applications on these arrays is handling a variable number of input channels. We consider the use of \acfp{GNN} as a solution to this challenge. We present a localization method using the Relation Network GNN, which we show shares many similarities to classical signal processing algorithms for \ac{SSL}. We apply our method for the task of \ac{SSL} and validate it experimentally using an unseen number of microphones. We test different feature extractors and show that our approach significantly outperforms classical baselines.
\end{abstract}

\input{sections/introduction}

\input{sections/prior_art}

\input{sections/method}
\input{sections/experimentation}
\input{sections/results_and_conclusion}

\clearpage
\bibliographystyle{IEEEtran}
\bibliography{sapstrings, Grinstein2023}

\end{document}

%% file: sapacronyms.txt
\acro{3GPP}{3rd Generation Partnership Project}
\acro{a-SLAM}[aSLAM]{Acoustic \aclu{SLAM}}
\acro{aSLAM}{Acoustic \aclu{SLAM}}
\acro{A-SNR}{A-weighted \acs{SNR}}
\acro{AAC}{Advanced Audio Coding\acroextra{. A lossy codec used for digital audio.}}
\acro{AAD}{Auditory Attention Detection}
\acro{AAS}{American Auditory Soc.}
\acro{AASP}{Audio and Acoustic Signal Processing}
\acro{ABC}{Analytical with or without Bias Compensation}
\acro{ABR}{Auditory-Brainstem Response}
\acro{ACAWD}{Archivable Core Actual-Word Database}
\acro{ACB}{Adaptive Codebook}
\acro{ACC}{Accuracy}
\acro{ACE}{Acoustic Characterization of Environments\acroextra{. A noisy reverberant speech corpus and IEEE challenge run by the SAP group at Imperial College}}
\acro{ACELP}{Algebraic Code-Excited Linear Prediction}
\acro{ACF}{Autocorrelation Function}
\acro{ACK}{Acknowledgement}
\acro{ACL}{Access Control List}
\acro{ACR}{Absolute Category Rating}
\acro{AD}{Audio Diarization}
\acro{ADC}{Analogue-to-Digital Converter}
\acro{ADM}{Adaptive Differential Microphone}
\acro{ADPCM}{Adaptive Differential Pulse Code Modulation}
\acro{ADSL}{Asymmetric Digital Subscriber Line}
\acro{AE}{Almost Everywhere}
\acro{AES}{Audio Engineering Society}
\acro{AES2}[AES]{Advanced Encryption Standard}
\acro{AGC}{Automatic Gain Control}
\acro{AH}{Amplitude Histogram}
\acro{AI}{Articulation Index}
\acro{AI2}[AI]{Artificial Intelligence}
\acro{AI3}[AI]{Audio Inpainting}
\acro{AIC}{Akaike Information Criterion}
\acro{AIFF}{Audio Interchange File Format}
\acro{AIR}{Acoustic Impulse Response}
\acro{AIR2}[AIR]{Aachen Impulse Response}
\acro{AIRD}{Aachen Impulse Response Database}
\acro{ALC}{Automatic Level Control}
\acro{ALCons}{Articulation Loss of Consonants}
\acro{AM}{Amplitude Modulation}
\acro{AMDF}{Average Magnitude Difference Function\acroextra{. A function with similar properties to the cross- or autocorrelation but that requires no multiplication to evaluate.}}
\acro{AMR}{Adaptive Multi-Rate}
\acro{AMR-NB}{Adaptive Multi-Rate Narrow Band}
\acro{AMR-WB}{Adaptive Multi-Rate Wide Band}
\acro{AMS}{Amplitude Modulation Spectrogram}
\acro{ANC}{Adaptive Noise Canceller}
\acro{ANS}{Autocorrelation-Based Noise Subtraction}
\acro{ANSI}{American National Standards Institute}
\acro{ANU}{Australian National University}
\acro{APLAWD}{Archivable Priority List Actual-Word Database}
\acro{AR}{Autoregressive}
\acro{ARD}{Arbeitsgemeinschaft der \"{o}ffentlich-rechtlichen Rundfunkanstalten der Bundesrepublik Deutschland\acroextra{. `Consortium (``Working group'') of the public-law broadcasting institutions of the Federal Republic of Germany'}}
\acro{ARMA}{Autoregressive Moving Average}
\acro{AS}{Audio Segmentation}
\acro{AS2}[AS]{Almost Surely}
\acro{ASA}{Acoustic Scene Analysis}
\acro{ASIO}{Audio Stream Input/Output\acroextra{. A computer soundcard protocol with low latency developed by Steinberg.}}
\acro{ASK}{Amplitude Shift Keying}
\acro{ASLP}{Audio, Speech, and Language Processing}
\acro{ASM}{Acoustic Scene Mapping}
\acro{ASR}{Automatic Speech Recognition}
\acro{ASS}{Approximate Spectrum Substitution}
\acro{AST}{Acoustic Source Tracking}
\acro{AST2}[AST]{Asymmetric Sampling in Time}
\acro{ATF}{Acoustic Transfer Function\acroextra{. The Fourier Transform of the \acs{RIR}.}}
\acro{ATLM}{Acoustic Tokenization and Language Modelling}
\acro{ATR}{Advanced Telecommunications Research Institute International\acroextra{, Kyoto, Japan}}
\acro{AUC}{Area under the curve}
\acro{AURORA}{Aurora Experimental Framework for the Evaluation of the Performance of Speech Recognition Systems under Noisy Conditions}
\acro{AUV}{Autonomous Underwater Vehicle}
\acro{AV}{Audio-Visual}
\acro{AWGN}{Additive White Gaussian Noise}
\acro{AWS}{Approximate Waveform Substitution}
\acro{BASIE}{Bayesian Adaptive Speech Intelligibility Estimation}
\acro{BCE}{Blind Channel Estimation}
\acro{BCL}{Bekesy Comfortable Loudness}
\acro{BCR}{Block-Coordinate Relaxation}
\acro{BEM}{Boundary Element Method}
\acro{BER}{Bit Error Rate}
\acro{BIBO}{Bounded-Input Bounded-Output}
\acro{BIC}{Bayesian Information Criterion}
\acro{Bk}{Berksons\acroextra{. A unit for measuring intelligibility.}}
\acro{BK}[B\&K]{Br{\"u}el and Kj{\ae}r}
\acro{BLSTM}{Bidirectional \acs{LSTM}}
\acro{BM}{Blocking Matrix}
\acro{BO-SCPHD}{Bearing-only \acs{SC-PHD}}
\acro{BO-SLAM}{Bearing-only \acs{SLAM}}
\acro{BOT}{Bearing-only tracking}
\acro{BP}{Basis Pursuit}
\acro{BPCC}{Basis Pursuit with Clipping Constraints}
\acro{BPDN}{Basis Pursuit Denoising}
\acro{BPM}{Beats Per Minute}
\acro{BPSK}{Binary Phase Shift Keying}
\acro{BR}{Barrodale and Roberts' (algorithm)}
\acro{BRI}{Basic Rate Index}
\acro{BSD}{Bark Spectral Distortion}
\acro{BSI}{Blind System Identification}
\acro{BSIM}{Binaural Speech Intelligbility Model}
\acro{BSS}{Blind Source Separation}
\acro{BSTOI}{Binaural \acs{STOI}}
\acro{BW}{Bandwidth}
\acro{BZ}{Back-to-Zero}
\acro{C4DM}{Centre for Digital Music}
\acro{C50}[$C_\textrm{50}$]{Clarity Index}
\acro{C-GFB}{Combination Gas-fired Boiler}
\acro{CART}{Classification and Regression Tree}
\acro{CASA}{Computational Auditory Scene Analysis}
\acro{CBR}{Constant Bit Rate}
\acro{CCC}{Cross-Correlation Coefficient}
\acro{CCCC}{DARPA CSR Corpus Coordinating Committee}
\acro{CCD}{Charge-Coupled Device}
\acro{CCI}{Call Clarity Index}
\acro{CCITT}{Consultative Committee for International Telephony and Telegraphy}
\acro{CCM}{Contralateral Competing Message}
\acro{CCR}{Comparison Category Rating}
\acro{CDB}{Constant Directivity Beamformer}
\acro{CDMA}{Code Division Multiple Access}
\acro{CELP}{Code-excited Linear Prediction}
\acro{CHIEF}{Combined Helmholtz Integral Equation Formulation}
\acro{CIT}{Constrained Initial Taps}
\acro{CL}{Clipping Level}
\acro{CLEAR}{Centre for Law Enforcement Audio Research}
\acro{CLID}{Cluster Identification Test}
\acro{CLT}{Central Limit Theorem}
\acro{CMA}{Constant Modulus Algorithm}
\acro{CMASI}{Coherence Modulated Acoustic Speckle Interferometry}
\acro{CMB}{Cosmic Microwave Background}
\acro{CNC}{Consonant-Nucleus-Consonant}
\acro{CNG}{Comfort Noise Generation}
\acro{CNN}{Convolutional Neural Network}
\acrodefplural{CNN}[CNNs]{Convolutional Neural Networks}
\acro{CODEC}{Coder-Decoder}
\acro{CPE}{Customer Premises Equipment}
\acro{CPHD}{Cardinalized \acs{PHD}}
\acro{CRACD}{Codec-robust Automatic Clipping Detector}
\acro{CRC}{Cyclic Redundancy Check}
\acro{CRNN}{Convolutional Recurrent Neural Network}
\acro{CS}{Channel Shortening}
\acro{CS2}[CS]{Compressive Sensing}
\acro{CSP}{Communications and Signal Processing}
\acro{CSR-WSJ}{Continuous Speech Recognition Wall Street Journal Phase 1\acroextra{ database}}
\acro{CST}{Connected Speech Test\acroextra{ speech corpus}}
\acro{CT}{Conversation Test}
\acro{CTTN}{Comparative Tolerance to Noise}
\acro{CV}{Coefficient of Variation}
\acro{CVNN}{Complex-Valued Neural Networks}
\acrodefplural{CV}{Coefficients of Variation}
\acro{CV2}[CV]{Constant Velocity}
\acro{CVC}{Consonant-Vowel-Consonant}
\acro{CW}{Continuous Wave}
\acro{CWM}{Centre-Weighted Median}
\acro{CWMY}{Centre-Weighted Myriad}
\acro{CWT}{Continuous Wavelet Transform}
\acro{DAC}{Digital-to-Analogue Converter}
\acro{DAM}{Diagnostic Acceptability Measure}
\acro{DAQ}{Data Acquisition}
\acro{DARPA}{Defense Advanced Research Projects Agency\acroextra{ of the United States Dept. of Defense}}
\acro{DARPA-RMD}{\acs{DARPA} 1000-Word Resource Management Database\acroextra{ for Continuous Speech Recognition}}
\acro{DAW}{Digital Audio Workstation}
\acro{dB}{Decibel}
\acro{dBFS}{\acs{dB} Full Scale}
\acro{DBN}{Deep Belief Network}
\acro{DBSTOI}{Deterministic \acs{BSTOI}}
\acro{DC}{Direct Current}
\acro{DCME}{Digital Circuit Multiplexing Equipment}
\acro{DCR}{Degradation Category Rating}
\acro{DCT}{Discrete Cosine Transform}
\acro{DDR}{Direct-to-diffuse ratio}
\acro{DDR3}{Double Data Rate Type Three}
\acro{DECT}{Digital European Cordless Telecommunication}
\acro{DeLILAH}{Detection of Clipping using Least Squares Residuals and Iterated Logarithm Amplitude Histogram}
\acro{DENBE}{\acs{DRR} Estimation using a Null-Steered Beamformer}
\acro{DET}{Detection Error Trade-off}
\acro{Dev}{Development\acroextra{ dataset of the \acs{ACE} Challenge}}
\acro{DFT}{Discrete Fourier Transform}
\acro{DI}{Directivity Index}
\acro{DI-NN}{Dual-Input Neural Network}
\acro{DirectX}{\acroextra{A programming interface developed by Microsoft for handling tasks related to multimedia.}}
\acro{DIRHA}{Distant-speech Interaction for Robust Home Applications\acroextra{ multi-microphone multi-language acoustic speech corpus}}
\acro{DMA}{Distributed Microphone Array}
\acro{DMA2}{Differential Microphone Array}
\acro{DMT}{Discrete Multi-Tone}
\acro{DMV}{Dynamically Managed Voice\acroextra{ system}}
\acro{DNN}{Deep Neural Network}
\acro{DNR}{Dynamic Noise Reduction}
\acro{DoA}{Direction-of-Arrival}
\acrodefplural{DoA}[DoAs]{Directions-of-Arrival}
\acro{DOA}{Direction-of-Arrival}
\acrodefplural{DOA}[DOAs]{Directions-of-Arrival}
\acro{DP}{Dynamic Programming}
\acro{DPCM}{Differential Pulse Code Modulation}
\acro{DPD}{Direct-Path Dominance}
\acro{DPD-MUSIC}{Direct-Path Dominance Multiple Signal Classification}
\acro{DR}{Douglas-Rachford}
\acro{DR2}{Dynamic Range}
\acro{DRM}{Diagnostic Rhyme Test}
\acro{DRR}{Direct-to-Reverberant Ratio}
\acro{DRNN}{Deep Recurrent Neural Net}
\acro{DRT}{Diagnostic Rhyme Test}
\acro{DSB}{Delay-and-Sum Beamformer}
\acro{DSOBM}{Deterministic \acs{SOBM}}
\acro{DSP}{Digital Signal Processing}
\acro{DSPS}{Double Sides Periodic Substitution}
\acro{DSR}{Distributed Speech Recognition}
\acro{DSWS}{Double Sides Waveform Substitution}
\acro{DTMF}{Dual Tone Multi-Frequency}
\acro{DTX}{Discontinued Transmission}
\acro{DWT}{Discrete Wavelet Transform}
\acro{EARS}{Embodied Audition for RobotS}
\acro{EBF}{Eigen-beamformer}
\acro{EBU}{European Broadcasting Union}
\acro{EC}{Echo Canceller}
\acro{EDC}{Energy Decay Curve}
\acro{EDR}{Energy Decay Relief}
\acro{EDF}{Energy Decay Function}
\acro{EEG}{Electroencephalography}
\acro{EER}{Equal Error Rate}
\acro{EFICA}{Efficient Fast Independent Component Analysis}
\acro{EF-NN}{Early Fusion Neural Network}
\acro{EIR}{Equalized Impulse Response}
\acro{EKF}{Extended Kalman Filter}
\acro{EKF-SLAM}[EKF-SLAM]{\acs{EKF} \acs{SLAM}}
\acro{EKF-SLAM2}[EKF-SLAM]{Extended Kalman Filter \acs{SLAM}}
\acro{EL}{Echo Loss}
\acro{ELF}{Extremely Low Frequency}
\acro{ELRA}{European Languages Research Association}
\acro{EM}{Estimation-Maximization\acroextra{. An iterative technique to solve certain optimization problems.}}
\acro{EMIB}{Eigenmike Microphone Interface Box}
\acro{ENF}{Electrical Network Frequency}
\acro{EPSRC}{Engineering and Physical Sciences Research Council}
\acro{EQ}{Equalisation}
\acro{ERB}{Equivalent Rectangular Bandwidth}
\acro{ERP}{Ear Reference Point (cf. ITU-T Rec. P.64 1999)}
\acro{ESA}{Early Stage Assessment}
\acro{ESM}{Equivalent Source Method}
\acro{ESPRIT}{Estimation of Signal Parameters via Rotational Invariance Techniques}
\acro{ETAN}{Equivalent Tolerance to Additional Noise} 
\acro{ETSI}{European Telecommunications Standards Institute}
\acro{EURASIP}{European Association for Signal Processing}
\acro{EUSIPCO}{European Signal Processing Conference}
\acro{Eval}{Evaluation\acroextra{ dataset of the \acs{ACE} Challenge}}
\acro{F1}{F1 Score}
\acro{FAR}{False Alarm Rate}
\acro{FastSLAM}[FastSLAM]{FActored Solution To Simultaneous Localization and Mapping}
\acro{FastSLAM2}[FastSLAM]{FActored Solution To \acs{SLAM}} 
\acro{FAU}{Friedrich-Alexander-Universit{\"a}t}
\acro{FB}{Forward-Backward}
\acro{FB2}[FB]{Fullband}
\acro{FBF}{Fixed Beamformer}
\acro{FBSS}{Forward-Backward Spatial Smoothing}
\acro{FCC}{Federal Communications Commission}
\acro{FC-NN}{Fully Connected Neural Network}
\acro{FDM}{Frequency Division Multiplexing}
\acro{FDR}{False Discovery Rate}
\acro{FDR2}[FDR]{Free-Decay Region}
\acro{FEC}{Forward Error Correction}
\acro{FEM}{Finite Element Method}
\acro{FFI}{Norwegian Defence Research Establishment}
\acro{FFT}{Fast Fourier Transform}
\acroindefinite{FFT}{an}{a}
\acro{FIFO}{First-In First-Out}
\acro{FIR}{Finite Impulse Response\acroextra{. A filter whose output is a weighted sum of past input values and whose system function contains only zeros and no poles.}}
\acro{FISM}{Fast Image Source Method}
\acro{FISST}{Finite Set STatistics}
\acro{FLOM}{Fractional Lower-Order Moments}
\acro{FLOS}{Fractional Lower-Order Statistics}
\acro{FM}{Frequency Modulation}
\acro{FMM}{Fast Multipole Method}
\acro{FN}{False Negative}
\acro{FN2}{Nth Formant}
\acro{FNR}{False Negative Rate}
\acro{FORTRAN}{The IBM Mathematical Formula Translating System}
\acro{FoV}{Field of View}
\acro{FP}{False Positive}
\acro{FPR}{False Positive Rate}
\acro{FPS}{Frames Per Second}
\acro{FRI}{Finite Rate of Innovation}
\acro{FSB}{Filter-and-Sum Beamformer}
\acro{FSK}{Frequency Shift Keying}
\acro{FT}{Flat-Top}
\acro{FWER}{Familywise Error Rate}
\acro{FWSSNR}{Frequency-Weighted Segmental \acs{SNR}}
\acro{FWSSRR}{Frequency-Weighted \acs{SSRR}}
\acro{G.711}{\acs{PCM} of Voice Frequencies}
\acro{GARCH}{Generalized Auto-regressive Conditional Heteroscedasticity}
\acro{GBW}{Gain Bandwidth Product}
\acro{GCC}{Generalized Cross-Correlation}
\acro{GCC-PHAT}{Generalized Cross-Correlation with Phase Transform\acroextra{ method of estimating \acs{TDoA}}}
\acro{GCF}{Global Coherence Field}
\acro{GCN}{Graph Convolutional Network}
\acro{GGD}{Generalized Gaussian Distribution}
\acro{GGD2}[G$\Gamma$D]{Generalised Gamma Distribution}
\acro{GM}{Gaussian Mixture}
\acro{GM-PHD}{Gaussian Mixture \acs{PHD}}
\acro{GMCA}{Generalized Morphological Component Analysis}
\acro{GMM}{Gaussian Mixture Model\acroextra{. An approximation to an arbitrary probability density function that consists of a weighted sum of Gaussian distributions}}
\acro{GNN}{Graph Neural Network}
\acro{GNSS}{Global Navigation Satellite System}
\acro{GPRS}{General Packet Radio Services}
\acro{GPS}{Global Positioning System}
\acro{GSC}{Generalized Sidelobe Canceller}
\acro{GSM}{Global System for Mobile Communications}
\acro{GSM-EFR}{\acs{GSM} Enhanced Full Rate Codec}
\acro{GSM-FR}{\acs{GSM} Full Rate Codec}
\acro{GSM-HR}{\acs{GSM} Half Rate Codec}
\acro{GUI}{Graphical User Interface}
\acro{HAAC}{High Amplitude Audio Capture}
\acro{HATS}{Head and Torso Simulator}
\acro{HERB}{Harmonicity-based dEReverBeration}
\acro{HFT}{Hands-Free Terminal}
\acro{HI}{Hearing-Impaired}
\acro{HIE}{Helmholtz Integral Equation}
\acro{HISAS}{High resolution Interferometric \ac{SAS}}
\acro{HINT}{Hearing-in-Noise Test}
\acro{HLT}{Human Language Technology}
\acro{HMM}{Hidden Markov Model}
\acro{HOS}{Higher-Order Statistics}
\acro{HPF}{High-Pass Filter}
\acro{HR}{Half Rate (\acs{GSM} Codec)}
\acro{HRI}{Human-Robot Interaction}
\acro{HRTF}{Head-related Transfer Function}
\acro{HSA}{Hearing, Speech, Audio\acroextra{ technology group at Fraunhofer IDMT}}
\acro{HSD}{Hybrid Steepest Descent}
\acro{HT}{Hannan-Thomson}
\acro{HTK}{Hidden Markov Model Tool Kit}
\acro{IBM}{Ideal Binary Mask}
\acro{IC}{Interference Canceller}
\acro{ICA}{Independent Component Analysis}
\acro{ICASSP}{Intl. Conf. on Acoustics, Speech and Signal Processing}
\acro{ID}{Identifier}
\acro{IDMT}{Institute for Digital Media Technology}
\acro{iDEN}{Integrated Digital Enhanced Network}
\acro{IEC}{International Electrotechnical Commission}
\acro{IEEE}{Institute of Electrical and Electronics Engineers}
\acro{IET}{Institute of Engineering and Technology}
\acro{IETF}{Internet Engineering Task Force}
\acro{IFFT}{Inverse Fast Fourier Transform}
\acro{IHC}{Inner Hair Cell}
\acro{IHT}{Iterative Hard Thresholding}
\acro{IID}{Independent and Identically Distributed}
\acro{IIR}{Infinite Impulse Response\acroextra{. A filter whose output is a weighted sum of both past input and past output values and whose system function contains both poles and zeros.}}
\acro{IL}{Iterated Logarithm\acroextra{, the logarithm of the logarithm}}
\acro{ILAH}{Iterated Logarithm Amplitude Histogram\acroextra{ clipping detection method}}
\acro{ILD}{Interaural Level Difference}
\acro{IMCRA}{Improved Minima Controlled Recursive Averaging\acroextra{. A technique for blindly estimating the spectrum of additive noise in a signal.}}
\acro{IMD}{Inter-Modulation Distortion}
\acro{IMSI}{International Mobile Subscriber Identity}
\acro{IMU}{Inertial Measurement Unit}
\acro{INMD}{In-service Non-intrusive Measurement Device}
\acro{INTERSPEECH}{Annual Conference of the \acs{ISCA}}
\acro{IO}{Infinitely Often}
\acro{IP}{Internet Protocol}
\acro{IPA}{International Phonetic Association}
\acro{IPD}{Interaural Phase Difference}
\acro{IRM}{Ideal Ratio Mask}
\acro{IRS}{Inverse repeated Sequence\acroextra{. A pseudo random sequence used for impulse response measurement.}}
\acro{IRS2}[IRS]{Intermediate Reference System}
\acro{IS}{Importance Sampling}
\acro{ISCA}{International Speech Communication Association}
\acro{ISDN}{Integrated Services Digital Network}
\acro{ISFT}{Inverse \acl{SFT}}
\acro{ISO}{Intl. Organization for Standardization}
\acro{ISP}{Intensity Spectral Profile}
\acro{IST}{Iterative Soft Thresholding}
\acro{ISTFT}{Inverse Short Time Fourier Transform}
\acro{ISVR}{Institute of Sound and Vibration Research\acroextra{, Southampton University, UK}}
\acro{ITD}{Interaural Time Difference}
\acro{ITF}{Interaural Transfer Function}
\acro{ITU}{International Telecommunication Union}
\acro{ITUR}[ITU-R]{International Telecommunication Union\acroextra{ Radiocommunication Sector}}
\acro{ITUT}[ITU-T]{International Telecommunication Union\acroextra{ Telecommunication Standardisation Sector}}
\acro{IUWT}{Isotropic Undecimated Wavelet (Starlet) Transform}
\acro{IWAENC}{Intl. Workshop on Acoustic Signal Enhancement}
\acro{IWAENC_PRE_2012}[IWAENC]{Intl. Workshop Acoustic Echo and Noise Control}
\acro{IWASE}{Intl. Workshop on Acoustic Signal Enhancement}
\acro{JADE}{Joint Approximate Diagonalization of Eigen-Matrices}
\acro{JPDA}{Joint Probabilistic Data Association}
\acro{JPEG}{Joint Photographic Experts Group}
\acro{KDE}{Kernel Density Estimate}
\acro{KF}{Kalman Filter}
\acro{KFD}{Kernel Fisher Discriminant\acroextra{ Analysis}}
\acro{KL}{Karhunen-Lo{\'{e}}ve}
\acro{KLT}{Karhunen-Lo{\'{e}}ve Transform}
\acro{KST}{Kolmogorov-Smirnov Test}
\acro{LAD}{Least Absolute Deviation}
\acro{LAN}{Local Area Network}
\acro{LARS}{Least Angle Regression}
\acro{LAT}[L$_{\textrm{AT}}$]{Equivalent Continuous Sound Level\acroextra{. Also called Leq}}
\acro{LBR}{Low Bitrate Redundancy}
\acro{LC}{Local Criterion}
\acro{LCMP}{Linearly Constrained Minimum Power}
\acro{LCMV}{Linearly Constrained Minimum Variance}
\acro{LCWM}{Linear Combination of Weighted Medians}
\acro{LDC}{Linguistic Data Consortium}
\acro{LEM}{Loudspeaker-Enclosure-Microphone System}
\acro{Leq}[L$_{\textrm{eq}}$]{Equivalent Continuous Sound Level\acroextra{. Also called LAT}}
\acro{LF}{Liljencrants-Fant\acroextra{. The developers of a glottal waveform model}}
\acro{LHS}{Left-Hand Side}
\acro{LID}{Language Identification}
\acro{LILAH}{\acs{LSR2}-\acs{ILAH}\acroextra{ clipping detection method}}
\acro{LIME}{LInear Predictive Multi-input Equalization\acroextra{ algorithm}}
\acro{LiNoPS}{Lightweight Noise Protection System}
\acro{LLN}{Law of Large Numbers}
\acro{LLR}{Log-Likelihood Ratio}
\acro{LLS}{Logarithmic Least Squares}
\acro{LMA}{Least Mean Absolute}
\acro{LMS}{Least Mean Squares\acroextra{ adaptive filter}}
\acro{LNA}{Low Noise Amplifier}
\acro{LoS}{Line-of-Sight}
\acro{LOT}{Listening-Only Test}
\acro{LP}{Linear Parameter}
\acro{LP2}[LP]{Linear Predictive}
\acro{LP3}[LP]{Linear Prediction}
\acro{LPC}{Linear Predictive Coding\acroextra{. An autoregressive model of speech production.}}
\acro{LQO}{Listening Quality Objective}
\acro{LREC}{Conf. on Language Resources and Evaluation}
\acro{LS}{Least-Squares}
\acro{LSA}{Log Spectral Amplitude}
\acro{LSB}{Lower Side-Band}
\acro{LSB2}[LSB]{Least Significant Bit}
\acro{LSD}{Log Spectral Distortion}
\acro{LSP}{Line Spectrum Pairs}
\acro{LSR}{Late Stage Review}
\acro{LSR2}[LSR]{Least Squares Residuals\acroextra{ clipping detection method}}
\acro{LSRT}{Least Squares Residuals with Thresholding\acroextra{ clipping detection method}}
\acro{LSTM}{Long Short-Term Memory}
\acro{LTASS}{Long Term Average Speech Spectrum}
\acro{LTI}{Linear Time Invariant}
\acro{LTP}{Long Term Prediction}
\acro{LU}{Loudness Unit}
\acro{LUFS}{Loudness Units Full-Scale}
\acro{MA}{Moving Average}
\acro{MAC}{Multiply Accumulate Operation}
\acro{MAD}{Median Absolute Deviation}
\acro{MAE}{Mean Absolute Error}
\acro{MAP}{Maximum \emph{a posteriori}}
\acro{MARDY}{Multichannel Acoustic Reverberation Database at York}
\acro{MARS}{Multivariate Adaptive Regression Splines}
\acro{MBF}{Matched Filter Beamformer}
\acro{MC}{Monte Carlo}
\acro{MCA}{Morphological Component Analysis}
\acro{MCC}{Matthew's Correlation Coefficient}
\acro{MCEQ}{MultiChannel EQualisation}
\acro{MCS}{Multidimensional Colouration Space}
\acro{MDCT}{Modified Discrete Cosine Transform}
\acro{MDL}{Minimum Description Length}
\acro{MDS}{Multidimensional Scaling}
\acro{Mel}{\acroextra{A non-uniform frequency scale corresponding to perceived frequency. It is approximately linear at low frequencies and logarithmic at high frequencies.}}
\acro{MELP}{Mixed Excitation Linear Prediction}
\acro{MFCC}{Mel-frequency Cepstral Coefficients}
\acro{MFS}{Method of Fundamental Solutions}
\acro{MFSK}{Multi-Frequency Shift Keying}
\acro{MHT}{Multi-Hypotheses Tracking}
\acro{MI}{Mutual Information}
\acro{MIMO}{Multiple-Input-Multiple-Output}
\acro{MINT}{Multiple-input/output INverse Theorem}
\acro{MIRS}{Motorola Integrated Radio System}
\acro{MIT}{Massachusetts Institute of Technology}
\acro{MIT-LCS}{Massacchusetts Institute of Technology Laboratory for Computer Science}
\acro{ML}{Maximum Likelihood}
\acro{MLD}{Masking Level Difference}
\acro{MLE}{Maximum Likelihood Estimation}
\acro{ML-TDoA}{Maximum Likelihood Time Difference of Arrival}
\acro{MLMF}{Machine Learning with Multiple Features}
\acro{MLP}{Multi-layer Perceptron}
\acro{MLS}{Maximum Length Sequence\acroextra{ of pseudo random bits.}}
\acro{MMSE}{Minimum Mean Squared Error}
\acro{MMT}{Multiscale Median Transform}
\acro{MNRU}{Modulated Noise Reference Unit}
\acro{MOM}{Mean of Maximum}
\acro{MOS}{Mean Opinion Score}
\acro{MOS-LQO}{Mean Opinion Score - Listening Quality Objective}
\acro{MP}{Matching Pursuit}
\acro{MP3}{\acs{MPEG}-2 Audio Layer III}
\acro{MPEG}{Moving Picture Experts Group}
\acro{MRF}{Markov Random Field}
\acro{MRP}{Mouth Reference Point (cf. ITU-T Rec. P.64 1999)}
\acro{MRT}{Modified Rhyme Test}
\acro{MS}{Minimum Statistics}
\acro{MSB}{Most Significant Bit}
\acro{MSC}{Mean Square Coherence}
\acro{MSE}{Mean Square Error}
\acro{MSN}{Multiple Subscriber Number}
\acro{MSNR}{Maximum \acs{SNR}}
\acro{MTF}{Modulation Transfer Function}
\acro{MTM}{Modified Trimmed Mean}
\acro{MUSCLE}{MeasUred Single-CLustEr}
\acro{MUSHRA}{Multi-stimuli Test with Hidden Reference and Anchor}
\acro{MUSHRAR}{Multi-stimuli Test with Hidden Reference and Anchor for Reverberation}
\acro{MUSIC}{Multiple Signal Classification}
\acro{MVDR}{Minimum Variance Distortionless Response\acroextra{ beamformer}}
\acro{MWF}{Multi-channel Wiener Filter}
\acro{NAH}{Nearfield Acoustic Holography}
\acro{NB}{Narrowband}
\acro{NCM}{Normalized Coherence Metric}
\acro{NH}{Normal-Hearing}
\acro{NIRA}{Non-Intrusive Room Acoustics}
\acro{NISE}{Non-Intrusive \acs{SNR} estimation}
\acro{NISI}{Non-Intrusive Speech Intelligibility Estimation}
\acro{NISQ}{Non-Intrusive Speech Quality Estimation}
\acro{NIST}{National Institute of Standards and Technology}
\acro{NL}{Noise Level}
\acro{NLA}{Non-Linear Approximation}
\acro{NLMS}{Normalized Least Mean Squares\acroextra{ adaptive filter}}
\acro{NMCFLMS}{Normalized Multichannel Frequency Domain Least Mean Square}
\acro{NMF}{Non-negative Matrix Factorization}
\acro{NOISEX-92}{Database to Study the Effect of Additive Noise on Speech Recognition Systems}
\acro{NOIZEUS}{Noisy Speech Corpus for Evaluation of Speech Enhancement Algorithms}
\acro{NOSRMR}{Normalized Overall \acs{SRMR}}
\acro{NOS}{Number of Sources}
\acro{NOSRMR}{Normalized Overall \acs{SRMR}}
\acro{NPM}{Normalized Projection Misalignment}
\acro{NPV}{Negative Predictive Value}
\acro{NR}{Noise Reduction}
\acro{NS}{Noise Suppression}
\acro{NSRMR}{Normalised \acs{SRMR}}
\acro{NSRR}{Normalized Signal-to-Reverberation Ratio}
\acro{NSRMR}{Normalised \acs{SRMR}}
\acro{NSV}{Negative-Side Variance}
\acro{NTP}{Network Time Protocol}
\acro{NV}{Noise-Vocoding}
\acro{OBL}{Octave Band Level}
\acro{ODF}{Overdrive Factor}
\acro{Ofcom}{Office of Communications\acroextra{, the independent regulator and competition authority for the UK communications industries}}
\acro{OFDM}{Orthogonal Frequency Division Multiplexing}
\acro{OHC}{Outer Hair Cell}
\acro{OIM}{Objective Intelligibility Measure}
\acro{OLA}{Overlap-add}
\acro{OM-LSA}{Optimally Modified Log-Spectral Amplitude{ Estimator}}
\acro{OMP}{Orthogonal Matching Pursuit}
\acro{OSI}{Open Systems Interconnection}
\acro{OSPA}{Optimal Subpattern Assignment}
\acro{OSRMR}{Overall \acs{SRMR}}
\acro{PAB-SRMR}{Per acoustic band \acs{SRMR}}
\acro{PALM}{Passive Acoustic Localization and Mapping}
\acro{PAMS}{Perceptual Analysis Measurement System}
\acro{PARCOR}{Partial Correlation Coefficients}
\acro{PB}{Phonetically Balanced}
\acro{PBF}{Positive Boolean Function}
\acro{PCA}{Principal Components Analysis}
\acro{PCM}{Pulse-Code Modulation}
\acro{PDA}{Personal Digital Assistant}
\acro{PDA2}[PDA]{Probabilistic Data Association}
\acro{PDE}{Partial Differential Equation}
\acro{pdf}{Probability Density Function}
\acro{PDF}{Probability Density Function}
\acro{PE}{Parameter Estimation}
\acro{PEASS}{Perceptual Evaluation for Audio Source Separation}
\acro{PEFAC}{Pitch Estimation Filter with Amplitude Compression}
\acro{PESQ}{Perceptual Evaluation of Speech Quality}
\acro{PF}{Psychometric Function}
\acro{pgfl}[p.g.fl.]{Probability Generating Functional}
\acro{PHAT}{Phase Transform}
\acro{PHD}{Probability Hypothesis Density}
\acro{PIP}{Peak-Image Pairing}
\acro{PIV}{Pseudo-Intensity Vector}
\acro{PL}{Pseudo-likelihood}
\acro{PLC}{Packet Loss Concealment}
\acro{PLL}{Phase Locked Loop}
\acro{PLP}{Perceptual Linear Prediction}
\acro{PLR}{Perceived Level of Reverberation}
\acro{PM}{Phase Modulation}
\acro{PMF}{Probability Mass Function}
\acro{PMOS}{Predicted Mean Opinion Score}
\acro{PNN}{Parameterized Neural Network}
\acro{POLQA}{Perceptual Objective Listening Quality Analysis}
\acro{POTS}{Plain Old Telephone Service}
\acro{PPP}{Poisson Point Process}
\acro{PPS}{Pulse-Per-Second}
\acro{PPV}{Positive Predictive Value}
\acro{PReLU}{Parametric Rectified Linear Unit}
\acro{PRLM}{Phoneme Recognition and Language Modelling}
\acro{PRP}{Pair-wise Relative Phase-ratio}
\acro{PSD}{Power Spectral Density}
\acro{PSK}{Phase Shift Keying}
\acro{PSNR}{Peak Signal-to-Noise Ratio}
\acro{PSOLA}{Pitch Synchronous Overlap Add\acroextra{. A method of scaling a signal in time and pitch independently.}}
\acro{PSQM}{Perceptual Speech Quality Measurement}
\acro{PSSL}{Positional Sound Source Localization}
\acro{PSTN}{Public Switched Telephone Network}
\acro{PWD}{Plane-Wave Decomposition}
\acro{PTA}{Pure-Tone Audiology}
\acro{QAM}{Quadrature Amplitude Modulation}
\acro{QILAH}{Quadrisected Iterated Logarithm Amplitude Histogram\acroextra{ clipping detection method}}
\acro{QMF}{Quadrature Mirror Filter}
\acro{QoE}{Quality-of-Experience}
\acro{QoS}{Quality-of-Service}
\acro{QPSK}{Quadrature Phase Shift Keying}
\acro{RADAR}{RAdio Detection And Ranging}
\acro{RASTA}{Relative Spectral}
\acro{RASTA-PLP}{Relative Spectral Perceptual Linear Prediction}
\acro{RASTI}{Room Acoustics Speech Transmission Index\acroextra{ (superseded by STIPA)}}
\acro{RBM}{Restricted Boltzmann Machine}
\acro{RB-PHD}{Rao-Blackwellised \acs{PHD}}
\acro{RBPF}{Rao-Blackwellised Particle Filter}
\acro{RC}{Relative Criterion}
\acro{RDT}[$R_\textrm{DT}$]{Reverberation Decay Tail}
\acro{RDTF}{Relative Direct Transfer Function}
\acro{ReLU}{Rectified Linear Unit}
\acro{RF}{Radio Frequency}
\acro{RFI}{Radio Frequency Interference}
\acro{RFS}{Random Finite Set}
\acro{RHS}{Right-Hand Side}
\acro{RIP}{Restricted Isometry Property}
\acro{RIR}{Room Impulse Response}
\acro{RLS}{Recursive Least Squares\acroextra{ adaptive filter}}
\acro{RLSD}{Relative Log Spectral Distortion}
\acro{RMCLS}{Relaxed Multichannel Least Squares}
\acro{RMCLS-CIT}{Relaxed MultiChannel Least-Squares with Constrained Initial Taps}
\acro{RMS}{Root Mean Square}
\acro{RMSE}{Root Mean Square Error}
\acro{RNN}{Recurrent Neural Network}
\acro{RelNet}{Relation network}
\acro{ROC}{Receiver Operating Characteristic}
\acro{ROHC}{Robust Header Compression}
\acro{RP}{Received Pronunciation}
\acro{RPE}{Regular Pulse Excitation}
\acro{RRTF}{Relative Real-Time Factor}
\acro{RS}{Reverberation Suppression}
\acro{RSM}{Reflector Source Method}
\acro{RSS}{Received Signal Strength}
\acro{RSV}{Room Spectral Variance}
\acro{RT}{Reverberation Time}
\acro{RTAN}{Robustness to Additional Noise}
\acro{RTF}{Real-Time Factor}
\acro{RTF2}[RTF]{Room Transfer Function}
\acro{RTF3}[RTF]{Relative Transfer Function}
\acro{RV}{Random Variable}
\acro{RVP}{Recursive Vector Projection}
\acro{RWTH}{Rheinisch-Westf\"{a}lische Technische Hochschule}
\acro{S50}[$S_{50}$]{Intelligibility Function Gradient at the \ac{SRT}}
\acro{SA}{Spectral Amplitude}
\acro{SAA}{Synthetic Aperture Audio}
\acro{SAP}{Speech And Audio Processing}
\acro{SAR}{Speech-to-Artifact Ratio}
\acro{SAR2}[SAR]{Speaker Alternation Rate}
\acro{SAR3}[SAR]{Synthetic Aperture \ac{RADAR}}
\acro{SAS}{Synthetic Aperture \ac{SONAR}}
\acro{SB}{Subband}
\acro{SC-PHD}{Single Cluster \acs{PHD}}
\acro{SCAF}{Single Channel Adaptive Filter}
\acro{SCB}{Stochastic Codebook}
\acro{SCNR}{Single-Channel Noise Reduction}
\acro{SCOT}{Smoothed Coherence Transform}
\acro{SCNR}{Single-Channel Noise Reduction}
\acro{SC-PHD}{Single Cluster \acs{PHD}}
\acro{SCR}{Signal-to-Competition Ratio}
\acro{SCRIBE}{Spoken Corpus of British English}
\acro{SCT}{Speech Corruption Toolkit}
\acro{SCT2}{Short Conversation Test}
\acro{SD}{Semantic Differential}
\acro{SDB}{Superdirective Beamformer}
\acro{SDD}{Spectral Decay Distributions}
\acro{SDDMSB}{\acs{SDD} with Mel-spaced frequency bands}
\acro{SDDSA}{\acs{SDD} with Mel-spaced frequency bands and selective averaging}
\acro{SDDSA-G}{\acs{SDDSA} with Gerkmann noise estimator}
\acro{SDDSA-H}{\acs{SDDSA} with Hendriks noise estimator}
\acro{SDR}{Software Defined Radio}
\acro{SDR2}{Speech}
\acro{SDRAM}{Synchronous Dynamic Random Access Memory}
\acro{SDT}{Speech Description Taxonomy}
\acro{SEDF}{Subband \ac{EDF}}
\acro{SELD}{Sound Event Localization and Detection}
\acro{SEMG}{Surface Electromyography}
\acro{SFDR}{Spurious Free Dynamic Range}
\acro{SFM}{Single Feature with Mapping}
\acro{SFT}{Spherical Fourier Transform}
\acro{SH}{Spherical Harmonic}
\acro{SHD}{Spectral Harmonic Decomposition}
\acro{SHD2}[SHD]{Spherical Harmonic Domain}
\acro{SIE}{System Identification Error}
\acro{SII}{Speech Intelligibility Index}
\acro{SIImod}{Speech Intelligibility Index in the modulation domain}
\acro{SIM}{Subscriber Identity Module}
\acro{SIMO}{Single-Input-Multiple-Output}
\acro{SINAD}{Signal-to-Noise and Distortion Ratio}
\acro{SIP}{Session Initiation Protocol}
\acro{SIR}{Signal-to-Interference Ratio}
\acro{SIR2}[SIR]{Sequential Importance Resampling}
\acro{SIREAC}{Simulation of REal Acoustics\acroextra{ Software Tool}}
\acro{SIS}{Sequential Importance Sampling}
\acro{SL}{Speech Level}
\acro{SLAM}{Simultaneous Localization and Mapping}
\acro{SLLN}{Strong Law of Large Numbers}
\acro{SLM}{Sound Level Meter}
\acro{SLF}{Spatial Likelihood Function}
\acro{SMA}{Spherical Microphone Array}
\acro{SMARD}{Single- and Multichannel Audio Recordings Database}
\acro{SMERSH}{Spatiotemporal Averaging Method for Enhancement of Reverberant Speech}
\acro{SMIR}{Spherical Microphone array Impulse Response}
\acro{SMPTE}{Society of Motion Picture and Television Engineers}
\acro{SMS}{Short Message Service}
\acro{SNR}{Signal-to-Noise Ratio}
\acro{SNR2}[SNR]{Speech-to-Noise Ratio}
\acro{SNT}{Subspace Noise Tracking\acroextra{ algorithm}}
\acro{SOBM}{STOI-optimal Binary Mask}
\acro{SONAR}{SOund Navigation And Ranging}
\acro{SOLA}{Synchronous Overlap Add\acroextra{. A method of scaling a signal in time and pitch independently.}}
\acro{SPC}{Specificity}
\acro{SPEECON}{Speech Databases for Consumer Devices}
\acro{SPHERE}{NIST SPeech Header REsources\acroextra{ software with embedded Shorten Compression}}
\acro{SPIN}{Speech Perception In Noise}
\acro{SPL}{Sound Pressure Level}
\acro{SPP}{Speech Presence Probability}
\acro{SPQA}{Speech Quality Assurance Package}
\acro{SQNR}{Signal-to-Quantization Noise Ratio}
\acro{SR}{Sparse Representation}
\acro{SR2}[SR]{Spectral Rotation}
\acro{SRA}{Statistical Room Acoustics}
\acro{SRI}{SRI International\acroextra{. Formerly Standford Research Institute}}
\acro{SRMR}{Speech-to-Reverberation Modulation Energy Ratio}
\acro{SRP}{Steered Response Power}
\acro{SRP-PHAT}{Steered Response Power with Phase Transform}
\acro{SRP-TDE}{Steered Response Power with Time Delay Estimation}
\acro{SRR}{Signal-to-Reverberation Ratio}
\acro{SRT}{Speech Reception Threshold\acroextra{ (also known as Speech Recognition Threshold)}}
\acro{SS}{Spectral Subtraction}
\acro{SSB}{Single Side-Band}
\acro{SSI}{Synthetic Sentence Identification}
\acro{SSL}{Sound Source Localization}
\acro{SSN2}[SSN]{Simultaneous Switching Noise}
\acro{SSN}{Speech-Shaped Noise}
\acro{SSNR}{Segmental \acs{SNR}}
\acro{SSOBM}{Stochastic \acs{SOBM}}
\acro{SSRR}{Segmental Signal-to-Reverberation Ratio}
\acro{SSW}{Staggered Spondaic Word}
\acro{STFT}{Short Time Fourier Transform}
\acro{STI}{Speech Transmission Index}
\acro{STIPA}{Speech Transmission Index for Public Address Systems}
\acro{STITEL}{Speech Transmission Index for Telecommunication Systems}
\acro{STMI}{Spectro-Temporal Modulation Index}
\acro{STNR}{\acs{NIST}'s Speech-to-Noise Ratio\acroextra{ Estimation Algorithm}}
\acro{STOI}{Short-Time Objective Intelligibility\acroextra{ measure}}
\acro{STQ}{Speech Processing, Transmission and Quality Aspects}
\acro{STSA}{Short Time Spectral Analysis}
\acro{STSA1}[STSA]{Short Time Spectral Amplitude}
\acro{SUS}{Semantically Unpredictable Sentences}
\acro{SVD}{Singular Value Decomposition}
\acro{SVM}{Support Vector Machine}
\acro{SWSOBM}{Stochastic \acs{WSOBM}}
\acro{T20}[$T_\textrm{20}$]{Reverberation Time\acroextra{ to decay by $20$ dB}}
\acro{T30}[$T_\textrm{30}$]{Reverberation Time\acroextra{ to decay by $30$ dB}}
\acro{T60}[$T_\textrm{60}$]{Reverberation Time\acroextra{ to decay by $60$ dB}}
\acro{TBM}{Target Binary Mask}
\acro{TDE}{Time Delay Estimation}
\acro{TDHS}{Time Domain Harmonic Scaling\acroextra{. A method of scaling a signal in time and pitch independently.}}
\acrodefplural{TDOA}[TDOAs]{Time-Differences-of-Arrival}
\acro{TDOA}{Time-Difference-of-Arrival}
\acrodefplural{TDOA}[TDOAs]{Time-Differences-of-Arrival}
\acro{TDT}{Tone Decay Test}
\acro{TF}{Time-Frequency}
\acro{TFS}{Temporal Fine Structure}
\acro{TFGM}{Time-Frequency Gain Modification\acroextra{. An approach to signal enhancement in which a signal is multiplied by a gain function in the time-frequency domain.}}
\acro{THD}{Total Harmonic Distortion}
\acro{TI}{Texas Instruments, Inc.}
\acro{TIMIT}{\acs{TI}-\acs{MIT} speech corpus}
\acro{TIPHON}{Telecommunication and Internet Protocol Harmonization Over Networks}
\acro{TLS}{Total Least-Squares}
\acro{TN}{True Negative}
\acro{TNR}{True Negative Rate}
\acro{TOA}{Time-of-Arrival}
\acrodefplural{TOA}[TOAs]{Times-of-Arrival}
\acro{TOF}{Time-of-Flight}
\acro{TOSQA}{Telekom Objective Speech Quality Assessmentt}
\acro{TP}{True Positive}
\acro{TP2}[TP]{Trivial Pursuit}
\acro{TPCC}{Trivial Pursuit with Clipping Constraints}
\acro{TPR}{True Positive Rate}
\acro{TSE}{Taylor Series Expansion}
\acro{TVAR}{Time-varying Autoregression}
\acro{UAV}{Unmanned Aerial Vehicle}
\acro{UDP}{User Datagram Protocol}
\acro{UFRJ}{Federal University of Rio de Janeiro}
\acro{UHF}{Ultra High Frequency}
\acro{UKF}{Unscented Kalman Filter}
\acro{ULA}{Uniform Linear Array}
\acro{ULF}{Ultra Low Frequency}
\acro{UMTS}{Universal Mobile Telecommunications Service}
\acro{US}{United States}
\acro{UTBM}{Universal Target Binary Mask}
\acro{VAD}{Voice Activity Detector}
\acro{VBR}{Variable Bit-Rate}
\acro{VCV}{Vowel-Consonant-Vowel}
\acro{VRD}{Variance of Decay-rates}
\acro{VGC}{Voice Grade Channel}
\acro{vMF}{von Mises-Fisher}
\acro{VoIP}{Voice Over Internet Protocol}
\acro{VRT}{Vlaamse Radio- en Televisieomroeporganisatie\acroextra{. (Flemish Radio and Television Broadcasting Organization)}}
\acro{VSELP}{Vector Sum-excited Linear Prediction}
\acro{VST}{Virtual Studio Technology\acroextra{. An interface standard developed by Steinberg for adding plugins to an audio editor.}}
\acro{WADA}{Waveform Amplitude Distribution Analysis}
\acro{WASN}{Wireless Acoustic Sensor Network}
\acrodefplural{WASN}[WASNs]{Wireless Acoustic Sensor Networks}
\acro{WASPAA}{Workshop on Applications of Signal Processing to Audio and Acoustics}
\acro{WAV}{Waveform Audio File Format}
\acro{WAVE}{Waveform Audio File Format}
\acro{WB}{Wideband}
\acro{WER}{Word Error Rate}
\acro{WGN}{White Gaussian Noise}
\acro{WLAN}{Wireless \acs{LAN}}
\acro{WLLN}{Weak Law of Large Numbers}
\acro{WM}{Working Memory}
\acro{WMA}{Windows Media Audio}
\acro{WNG}{White Noise Gain}
\acro{WSS}{Weighted Spectral Slope}
\acro{WSOBM}{Weighted \acs{SOBM}}
\acro{WSTOI}{Weighted \acs{STOI}}
\acro{ZOS}{Zero-Order Statistics}

%% file: sections/introduction.tex
\section{Introduction}
  \vspace{-0.3cm}
\acf{DMA} signal processing \cite{Bertrand2011} is an active field in the acoustic signal processing community, with important applications in speech enhancement, noise reduction and \acf{SSL} \cite{Tzirakis2021, Bertrand2011, Cobos2017a}. In contrast to centralized microphone arrays \cite{brandstein2001microphone}, \acp{DMA} may be created through the wireless connection of multiple distributed devices such as cell phones, laptops and virtual assistants. In this context, they are also frequently referred to as Ad-hoc microphone arrays, or \acfp{WASN}.

Although \acp{DMA} bring advantages in terms of acoustic coverage with respect to centralized arrays, they also bring challenges. One such challenge forms the focus of this paper, namely, having a dynamic number of input microphone channels, as a \ac{DMA} may be created using the devices present in a dynamic scene. This number may change in runtime due to many reasons, including software or hardware failures of individual devices, battery depletion, or the device being removed from the scene. This restricts the application of many of the deep learning methods that have been successfully applied to centralized microphone networks such as \cite{Adavanne2018, Chakrabarty2017}, which require a static input size. Conversely, classical \ac{SSL} approaches such as \cite{Aarabi2003} are able to function on an arbitrary number of microphones.

In this work, we propose the use of \acfp{GNN} \cite{Zhou2020, Battaglia2018, Santoro2017} as a suitable way of processing \ac{DMA} signals for the task of \ac{SSL}. We adopt a \ac{GNN} variant called the \acf{RelNet} \cite{Santoro2017}. We validate our approach for the task of localizing a single static source in multiple scenarios, showing it to outperform the baselines. The main contribution of this work is the first application of \acp{GNN} for the task of \ac{SSL}, allowing our method to handle a variable number of microphone channels. Furthermore, our approach can work on unseen microphone coordinates and room dimensions through a metadata fusion procedure.

This paper continues by providing a problem statement in \autoref{sec:statement}. \autoref{sec:prior-art} includes a review of related work on \acp{DMA} using deep learning, as well as a review of classical \ac{SSL} methods and the \ac{RelNet} \ac{GNN}, which serve as building blocks for our model. In \autoref{sec:method}, we describe our proposed approach. \autoref{sec:experimentation} describes our experimental validation, and \autoref{sec:results} presents the results and \autoref{sec:conclusion} concludes the paper.

\begin{figure}[h]
  \centering
  \includegraphics[scale=0.9]{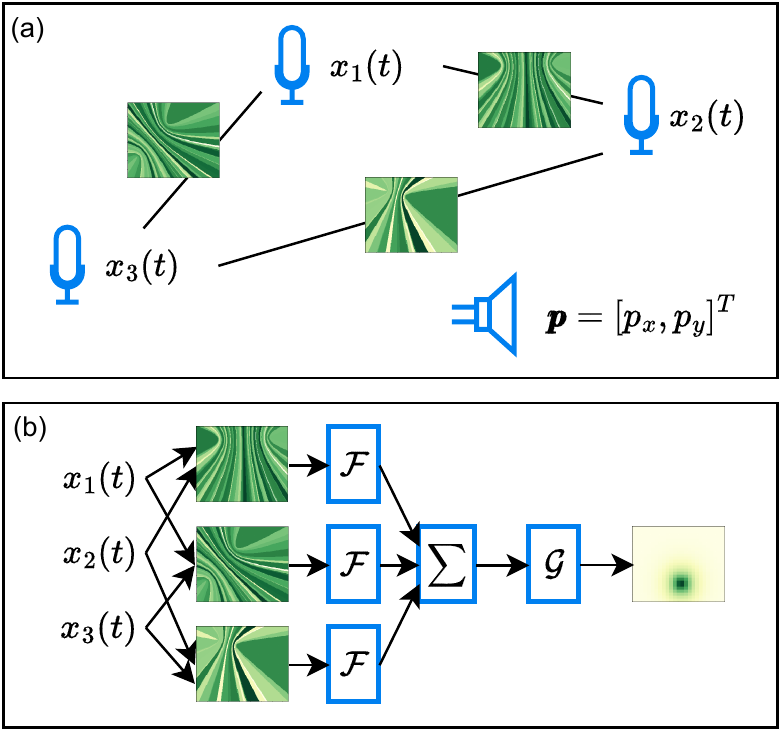}
  \vspace{-0.2cm}
  \caption{(a): Example of a graph of distributed microphones. (b): Representation of the GNN-SLF model for three microphones. The computation of the heatmaps is described in \autoref{sec:method}.}
  \label{fig:nets}
\end{figure}

\section{Problem statement} \label{sec:statement}
Our goal is to estimate the 2D coordinates $\hat{\pmb{p}}_s$ of a sound source located at $\pmb{p}_s = [p_s^x \, p_s^y]^T$ within a reverberant room of known dimensions $\pmb{d} = [d^{x} \, d^{y} \, d^{z}]^T$. The source emits a speech signal $s(t)$ at instant $t$. Besides the source, $M$ microphones are present in the room, where microphone $m$ has a known position $\pmb{p}_m = [p_m^x \, p_m^y \,  p_m^z]^T$, and receives a signal $x_m(t)$ modeled as
\begin{equation} \label{eq:received_signal}
    x_m(t) = a_m s(t - \tau_m) + \epsilon_m(t) ,
\end{equation}
where $a_m$ is a scaling factor representing the attenuation suffered by the wave propagating from $\pmb{p}_s$ to $\pmb{p}_m$. $\tau_m$ represents the time delay taken for a sound wave to propagate from the source to the microphone, and $\epsilon_m$ models the noise and reverberation. We assume $\tau_m$ to be equal to $c^{-1} \Vert \pmb{p}_m - \pmb{p}_s \Vert_2 $, the distance between the source and the microphone divided by the speed of sound $c$.

In our method and baselines, the microphone signals are sampled and processed in frames of size $L$, defined as $\pmb{x}_m(t) = [x_m(t - (L - 1)T_s) ... x_m(t)]^T$, where $T_s$ is the sample period. Finally, we also define a metadata vector $\pmb{\phi}$ as
\begin{equation} \label{eq:phi}
    \pmb{\phi} = [p_1^x \, p_1^y ... d^{y} \, d^{z}]^T,
\end{equation}
which serves as a secondary input to our method, allowing it to function on any room dimensions and microphone coordinates.

%% file: sections/prior_art.tex
\vspace{-0.2cm}
\section{Related work}
\label{sec:prior-art}

\subsection{Classical SSL methods}

Our proposed method can be seen as a generalization of classical grid-based \ac{SSL} methods such as the \ac{TDOA} \cite{So2011, Gustafsson2003a}, \ac{SLF} \cite{Aarabi2003, Pertila2008} and energy-based \cite{Li2003c} approaches. These approaches share many similarities, which are summarized by their shared behaviour described in \autoref{alg:ssl}.


\begin{algorithm} \caption{Classical SSL methods}
\label{alg:ssl}
\begin{algorithmic}
\Function{estimate\_source\_location}{$\pmb{X}, \pmb{\phi}$}
\State $ \pmb{u} \gets \pmb{0}$
\ForEach {$i \in [1..M]$}
    \ForEach {$j \in [(i+1)..M] $}
        \State $\pmb{u} \pluseq \mathcal{F}(\pmb{x}_i, \pmb{x}_j; \pmb{\phi}(i,j))$
    \EndFor
\EndFor
\State \Return {$\mathcal{G}(\pmb{u})$}
\EndFunction
\end{algorithmic}
\end{algorithm}

\autoref{alg:ssl} starts with the creation of an empty grid $\pmb{u}$, which we assume to be a flattened 2D for our applications. The next step consists of computing a \textit{relation} $\mathcal{F}$ between each pair of microphones $(i,j)$ available, using their signals $(\pmb{x}_i, \pmb{x}_j)$ as well as the \textit{metadata} available $\pmb{\phi}$, consisting of the microphone and room dimensions and the speed of sound. These relations consist of assigning, for each cell within the grid, a value expressing how likely a source is to be in a particular grid cell.

The relations between all pairs are aggregated through summation (or multiplication, see \cite{Pertila2008}) to generate a heatmap gathering all pairwise information. Depending on whether the problem is formulated using a \ac{LS} or \ac{ML} approach, the minimum or maximum value of the grid will respectively correspond to the location of the source \cite{So2011}. $\mathcal{G}$ is therefore a peak-picking function, whose goal is to select the grid cell where the source is located.

The \ac{TDOA}, \ac{SLF} and energy-based methods differ mainly by the function $\mathcal{F}$ computed. Each cell within the grid represents a candidate source location which has a theoretical \ac{TDOA} between the two microphones. In the \ac{TDOA} method, each grid cell is assigned the distance between its theoretical \ac{TDOA} and the measured \ac{TDOA}, computed by picking the peak of the generalized cross-correlation function between the microphones' signals, typically computed using the \ac{GCC-PHAT} \cite{Knapp1976b}.

In the \ac{SLF} method, each cell receives the cross-correlation value at the lag corresponding to its \ac{TDOA}. \ac{SLF} is shown to be equivalent to \ac{SRP} \cite{DiBiase2000}. Finally, the energy-based method uses a metric based on the ratio of the two microphone signals' energies. In \autoref{fig:nets}a, the edges of the graph represent maps computed using the \ac{SLF} method.

\vspace{-0.5cm}

\subsection{Neural network methods for DMA signal processing}

Classical \ac{SSL} methods normally do not account for room reverberation, which may divert the heatmap's peak from the true source location, or reduce its sharpness. Neural networks can become robust to reverberation if trained on suitable scenarios. Here we review works on neural networks for \acp{DMA}. 

In \cite{Furnon2021}, an attention-based neural network capable of handling connection failures is proposed for the task of speech enhancement. Unlike our method, this network is limited to a maximum number of input microphones channels. In \cite{Wang2020} and \cite{Luo2020}, variable-input processing is achieved through a global average pooling scheme. 

Two works have explored \acp{GNN} for acoustic signal processing. In \cite{Luo2022}, a \ac{GNN} is used to profile noise within a railway setting. However, their work the source signal to be known beforehand, limiting its application in many scenarios. This restriction is not present in our proposed approach. In \cite{Tzirakis2021}, a \ac{GCN} \cite{Kipf2017} is used in conjunction with an encoder-decoder network for the task of speech enhancement. Conversely, we do not use an encoder-decoder and explore the Relation Network GNN, which we show to be well suited for the task of \ac{SSL}.

\subsection{Relation Networks}

We choose the \acf{RelNet} \cite{Santoro2017} as our graph network architecture due its conceptual similarities to classical \ac{SSL} methods. \acp{RelNet} were introduced in the context of visual question answering. The input of the network consists of a set of \textit{nodes}, represented by feature vectors $\pmb{X} = \{\pmb{x}_1, \pmb{x}_2, ..., \pmb{x}_M\}$. The network $\mathcal{RN}$ may be summarized as

\vspace{-0.3cm}
\begin{equation} \label{eq:relation_net}
    \hat{\pmb{y}} = \mathcal{RN}(\pmb{X}) = \mathcal{G}\bigg(\sum_{i \neq j} \mathcal{F}(\pmb{x}_i, \pmb{x}_j)\bigg),
\end{equation}

where \eqref{eq:relation_net}, $\mathcal{F}$ generates a \textit{relation} between nodes $(i, j)$. These relations are summed together, and this sum is the input to $\mathcal{G}$, which produces the answer $\hat{\pmb{y}}$ to the target question. The nodes $\pmb{x}_i$ and the relations $\mathcal{F}(\pmb{x}_i, \pmb{x}_j)$ can be seen as a complete undirected graph $\pmb{G} = (\{\pmb{x}_i\}, \{\mathcal{F}(\pmb{x}_i, \pmb{x}_j)\})$. As in \cite{Santoro2017}, we implement both $\mathcal{F}$ and $\mathcal{G}$ as \acp{MLP}, trained jointly using backpropagation.

%% file: sections/method.tex
\vspace{-0.2cm}
\section{Method} \label{sec:method}

A diagram of our proposed network is shown in \autoref{fig:nets}b. Using a \ac{RelNet} allows our approach to first process pairs of microphone signals into features, and later combine them through summation. This allows it to function on a variable number of input microphones. Furthermore, our method can operate on unknown room dimensions and microphone coordinates by combining this metadata $\pmb{\phi}$ before estimating the source location.

The input to our method consists of the set of $M$ microphone signal frames $\{\pmb{x}_m\}$,
where $\pmb{x}_m$ is a vector of size $L$ representing a frame of recordings, and a metadata vector $\pmb{\phi}$ containing relevant information such as the microphone coordinates and room dimensions. We define the relation function $\mathcal{F}$ as
\begin{equation} \label{eq:relation_func}
    \mathcal{F}(\pmb{x}_i, \pmb{x}_j; \pmb{\phi}) = \text{MLP}(\mathcal{H}(\pmb{x}_i, \pmb{x}_j; \pmb{\phi})),
\end{equation}
Where $\text{MLP}$ is a multi-layer perceptron and $\mathcal{H}$ is a preprocessing or feature extraction function. The inclusion of a preprocessing function allows us to use the classical features such as \ac{GCC-PHAT} or \ac{SLF}. Conversely, post-processing these functions using a \ac{MLP} allows us to improve these features by introducing learned rules, as we will show for the application of \ac{SSL}.

In turn, the relation fusion function is chosen as $\mathcal{G}(\pmb{u}) = \text{MLP}(\pmb{u})$, where $\pmb{u}$ represents the sum of all pairs of relations as in \autoref{alg:ssl}. This function is a substitution of the peak-picking algorithm in \autoref{alg:ssl}, expanding its functionality for other possible applications.

As in \cite{Santoro2017}, we train the weights $\pmb{w}_\mathcal{F}$ and $\pmb{w}_\mathcal{G}$ of the MLPs in $\mathcal{F}$ and $\mathcal{G}$ jointly through a gradient-based procedure, by minimizing an application-specific loss function $\mathcal{L}(y, \hat{y})$ between the network output $\hat{y}$ and target $y$:
\vspace{-0.3cm}
\begin{equation} \label{eq:loss}
\begin{split}
    \pmb{w}_\mathcal{F} = \pmb{w}_\mathcal{F} - \lambda_\mathcal{F} \frac{\partial \mathcal{L}(\pmb{y}, \hat{\pmb{y}})}{\partial \pmb{w}_\mathcal{F}} & \\
    \pmb{w}_\mathcal{G} = \pmb{w}_\mathcal{G} - \lambda_\mathcal{G} \frac{\partial \mathcal{L}(\pmb{y}, \hat{\pmb{y}})}{\partial  \pmb{w}_\mathcal{G}},
\end{split}
\end{equation}

Where $(\lambda_\mathcal{F}, \lambda_\mathcal{G})$ are the learning rates, usually defined by the optimizer used, such as Adam \cite{Kingma2017}.

We experiment with two preprocessing functions $\mathcal{H}$ for our relation function $\mathcal{F}$. The first is the cross-correlation between the two microphones, computed using the \ac{GCC-PHAT} method. In this case, the network needs to learn to map time lags into space. As an alternative, we project the cross-correlation into space using the \ac{SLF} method. The output of this method is a flattened $N \times N$ grid or a $N^2$ vector. In this case, the network needs to learn to denoise the maps which may have been corrupted by reverberation.

A final step in the feature extraction step is concatenating the microphone coordinates of the pair as well as its room dimensions into the features. This is especially important for the GCC-PHAT feature extractor, as the network must learn how to project the temporal information into space.

The target of the \ac{MLP} of function $\mathcal{G}$ is to further enhance the summed maps produced by $\mathcal{F}$. Its output has the same size as $\mathcal{F}$, representing a flattened $N \times N$ grid of cells centered at coordinates $\{\pmb{p}_{u,v}\}$ within the room. The target value $y(u, v)$ of each grid cell $(u, v)$ is computed as
\begin{equation}
    y(u, v) = e^{- \Vert \pmb{p}_{u,v} - \pmb{p}_s \Vert_2},
\end{equation}
where $\pmb{p}_s$ is the target source location. Note the maximum value of 1 occurs when $\pmb{p}_{u,v} = \pmb{p_s}$ and approaches 0 exponentially as the distance between $\pmb{p}_{u,v}$ and $\pmb{p}_s$ increases. We use the mean absolute error between the network output and target as our loss function. This formulation allows for detection of multiple sources, which can be extracted through peak-picking. However, in this work, we focus on the detection of a single source.

%% file: sections/experimentation.tex
\vspace{-0.2cm}
\section{Experimentation} \label{sec:experimentation}

This section describes our experiments with our proposed network for \ac{SSL} described in the previous section. We refer to our proposed methods as GNN-GCC for the network using the GCC-PHAT feature extractor and GNN-SLF for the one using the SLF extractor. We compare our approach with two baselines, the classical \acf{TDOA}-based and \acf{SLF}-based approaches, as described in \autoref{sec:prior-art}. We provide a public repository containing all methods on Github \footnote{\url{https://github.com/egrinstein/gnn_ssl}}

\subsection{Dataset}
We test our approach using synthetically generated data using the image source method \cite{Allen1979b}, generated using the Pyroomacoustics library \cite{Scheibler2018a}. To demonstrate that our approach is able to operate with a different number of microphones than it was trained on, the training set for our \ac{GNN} uses training examples containing $\{5,7\}$ microphones, while the test set examples contain $\{4,5,6,7\}$ microphones.

For each dataset sample, we randomly select two numbers from a uniform distribution in the interval [3, 6]~m representing the room's width and length. The room's height is uniformly selected from the interval [2, 4]~m. The room' sreverberation time is sampled uniformly from the interval [0.3, 0.6]~s using Eyring's formula \cite{Neubauer2001}. We place the microphones and source randomly within the room, with the restriction of each device being at least 0.5~m from each other and the room's walls. Each source is set to play a speech sample from the VCTK corpus \cite{Yamagishi2019b}. The \ac{SNR} in each microphone is set at 30~dB, simulated by adding \ac{WGN} independently to each channel to the auralizations generated using the image source method. The training, validation and test datasets contain respectively 15,000, 5000 and 10,000 examples.

\vspace{-0.3cm}
\subsection{Method hyperparameters}
We train the networks for a maximum of 100 epochs with early stopping if the validation loss stops increasing after 3 epochs. We employ a learning rate of 0.0005 using the Adam optimizer \cite{Kingma2017}. We use a batch size of 32. These parameters were chosen empirically. All grids used are of dimensions $25 \times 25$. Our input frame size used is L=500~ms. For the GCC-PHAT method, we use a \ac{DFT} of $1,024$ samples. Since the maximum TDOA value is bounded by the room's diagonal, we only select the central 200 correlation bins, similar to \cite{He2018a}. In our proposed method, our relation function's MLP contains 3 layers, each of output size 625. The function $\mathcal{G}$'s MLP consists of 3 layers, all with an output size of 625 neurons. We use a ReLU activation function for all layers except for the output, which uses no activation.

The grids computed in the \ac{SLF} and \ac{TDOA} baselines as well as the feature extractor in the GNN-SLF method have a size of $25\times 25$. The source estimation procedure in the baselines and proposed methods consists of picking the location of the highest value in the \ac{SLF} method, and the lowest on in the \ac{SLF} method.

%% file: sections/results_and_conclusion.tex
\vspace{-0.3cm}
\section{Results} \label{sec:results}
\vspace{-0.6cm}

\begin{figure}[ht]
  \centering
  \includegraphics[scale=0.62]{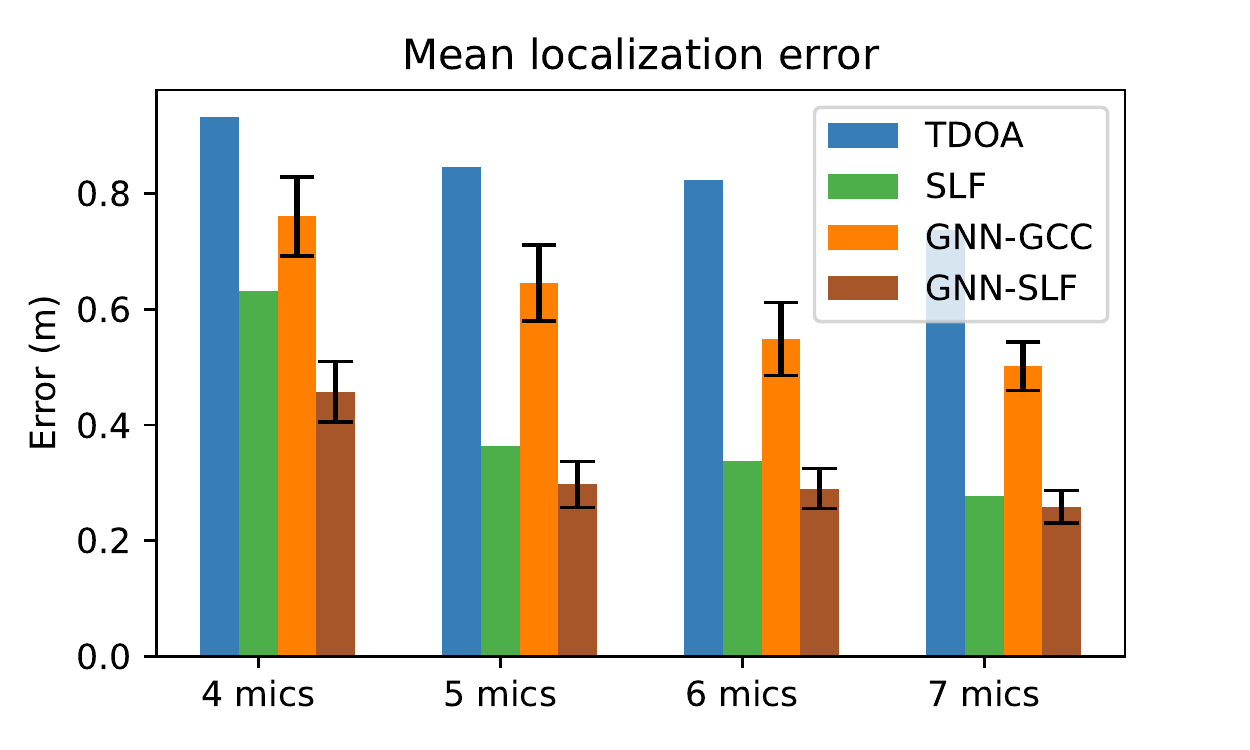}
  \vspace{-0.9cm}
  \caption{Localization error for our proposed methods and baselines.}
  \label{fig:errors}
\end{figure}

The metric used to evaluate the methods consists of the mean euclidean distance between the estimated and true source location on the test set. The results are shown in \autoref{fig:errors}. Note that although we test all methods on unseen simulations containing $\{4, 5, 6, 7\}$ microphones, our method was only trained using examples containing $\{5,7\}$ microphones. To ensure a fair comparison, the networks were trained multiple times. The black bars show their standard deviation.

We can see that the GNN-SLF method outperforms all others, demonstrating the effectiveness of the approach. The biggest relative improvement of 29\% with respect to classical SLF is observed for four microphones. An explanation is that when there are fewer measurements available improving or discarding them becomes crucial, which may be the operation being performed by the network. We also see that GNN-GCC performed poorly, only surpassing the TDOA baseline. This indicates that requiring the network to learn to map time delays to spatial position is a more demanding task than dealing with the already spatialized information. 

  \vspace{-0.3cm}
  
\section{Conclusion and future work} \label{sec:conclusion}
We applied the \ac{RelNet}, a type of \ac{GNN} for the task of \ac{SSL} on distributed microphone arrays. Our results show the \ac{RelNet} is able to significantly improve the localization performance over classical localization algorithms, achieving a 29\% improvement in the case of 4 microphones. We also show the method generalizing to an unseen number of microphones. Future directions include testing approach for localizing multiple sources and learning graph topologies different than the complete graph.